\documentclass[twocolumn,english,prb,groupedtaddress]{revtex4}
\usepackage[T1]{fontenc}
\usepackage[latin9]{inputenc}
\usepackage{babel}
\usepackage{caption}
\usepackage{float}
\usepackage[caption=false]{subfig}
\usepackage{verbatim}
\usepackage{amsmath}
\usepackage{amssymb}
\usepackage{graphicx}
\usepackage{esint}
\usepackage{graphicx}
\usepackage{amssymb}

\usepackage{hyperref}
\hypersetup{
   bookmarks=true,         
    unicode=false,          
    pdfmenubar=true,        
    pdffitwindow=true,      
    pdftitle={My title},    
    pdfauthor={Author},     
   pdfsubject={Subject},   
    pdfcreator={Creator},   
    pdfproducer={Producer}, 
    pdfkeywords={keywords}, 
    pdfnewwindow=true,      
   colorlinks=true,       
   linkcolor=red,          
    citecolor=blue,        
    filecolor=magenta,      
    urlcolor=cyan           
}

\makeatletter
\@ifundefined{textcolor}{}
{%
 \definecolor{BLACK}{gray}{0}
 \definecolor{WHITE}{gray}{1}
 \definecolor{RED}{rgb}{1,0,0}
 \definecolor{GREEN}{rgb}{0,1,0}
 \definecolor{BLUE}{rgb}{0,0,1}
 \definecolor{CYAN}{cmyk}{1,0,0,0}
 \definecolor{MAGENTA}{cmyk}{0,1,0,0}
 \definecolor{YELLOW}{cmyk}{0,0,1,0}
 }

\@ifundefined{definecolor}
 {\usepackage{color}}{}

\usepackage{times}\usepackage{mathrsfs}\usepackage{paralist}\usepackage{ifthen}\usepackage{dsfont}\@ifundefined{definecolor}{\usepackage{color}}{}

\def\be{\begin{equation}}
\def\ee{\end{equation}}
\def\bea{\begin{eqnarray}}
\def\eea{\end{eqnarray}}

\makeatother

\begin{document}

\title{Suppression of spontaneous currents in Sr$_2$RuO$_4$ by surface disorder  }
\author{Samuel Lederer$^1$, Wen Huang$^2$, Edward Taylor$^2$, Srinivas Raghu$^{1,3}$, and Catherine Kallin$^{2,4}$} 
\affiliation{$^1$Department of Physics, Stanford University, Stanford, California, 94305, USA}
\affiliation{$^2$Department of Physics and Astronomy, McMaster University, Hamilton, Ontario, L8S 4M1, Canada}
\affiliation{$^3$SLAC National Accelerator Laboratory, 2575 Sand Hill Road, Menlo Park, CA 94025}
\affiliation{$^4$Canadian Institute for Advanced Research, Toronto, Ontario M5G 1Z8, Canada}

\begin{abstract}
A major challenge to the chiral $p$-wave hypothesis for the pairing symmetry of the unconventional superconductor Sr$_2$RuO$_4$ is the null result of sensitive scanning magnetometry experiments designed to detect the expected spontaneous charge currents. Motivated by junction tunneling conductance measurements which indicate the quenching of superconductivity at the surfaces of even high-purity samples, we examine the spontaneous currents in a chiral $p$-wave superconductor near a normal metal / superconductor interface using the lattice Bogoliubov-de Gennes equations and Ginzburg-Landau theory, and find that the edge current is suppressed by more than an order of magnitude compared to previous estimates. These calculations demonstrate that interface details can have a quantitatively meaningful effect on the expectations for magnetometry experiments.

\end{abstract}

\maketitle
\section{Introduction}
Strontium Ruthenate, Sr$_2$RuO$_4$, is an unconventional superconductor ($T_c=1.5K$)\cite{Mackenzie2003} for which there exists substantial evidence for odd-parity pairing\cite{Ishida1998,Nelson2004,Kidwingira2006,Jang2011} as well as for the spontaneous breaking of time reversal symmetry below $T_c$\cite{Luke1998,Luke2000,Xia2006}. These observations lead naturally to the conclusion that the pairing symmetry is chiral $p$-wave ($p_x\pm i p_y$ ), a two dimensional analog of the A-phase of superfluid $^3$He.  Though this is the leading phenomenological hypothesis, it is seemingly contradicted by several experiments.  Prominent among these are high resolution scanning magnetometry measurements\cite{Kirtley2007,Hicks2010}, which image magnetic fields across several $\mu m$ of sample (including the sample edge) and see no sign of the expected spontaneous currents.

The presence of spontaneous, persistent charge currents at edges and domain walls is a robust consequence of time-reversal symmetry breaking superconductivity. However, the magnitude of these currents is determined by microscopic details -- they are neither quantized nor universal.  The reason that the null result of the scanning magnetometry experiments poses such a challenge to the chiral $p$-wave hypothesis is quantitative -- spontaneous currents of size comparable to theoretical estimates\cite{Matsumoto1999,Furusaki2001,Stone2004,Imai2012,Sauls2011} would give a magnetic signal more than two orders of magnitude greater than the experimental resolution. Magnetometry measurements on mesoscopic samples\cite{Jang2011} also see no signs of these currents.

In this paper we calculate the spontaneous surface currents for a family of models consistent with the phenomenology of superconductivity in Sr$_2$RuO$_4$. Motivated  by $a$-axis tunneling experiments\cite{Kashiwaya2011}, we employ a different interface condition than previous studies, modeling the surface region as a normal metal layer adjoining the superconducting bulk. We find that, compared to previous estimates, the expected magnetic signal from edge currents is reduced by over an order of magnitude.  These calculations demonstrate that interface details can have a quantitatively meaningful effect on the expectations for magnetometry experiments.

\section{Surface imperfection}
The assumption of specular surface scattering as employed in\cite{Matsumoto1999,Furusaki2001,Stone2004,Imai2012,Sauls2011} requires an atomically smooth surface. $ab$ faces of Sr$_2$RuO$_4$ can be cleaved, but $ac$ and $bc$ faces are typically polished to a smoothness of several $nm$\cite{Kirtley2007}, on the order of ten lattice constants. In $a$-axis junction tunneling conductance measurements, signatures of superconductivity at the surface are present only at the sub-$1\%$ level on top of a substantial smooth background\cite{Kashiwaya2011}, as shown in Fig. 2 of that reference.  Accordingly, the best indication from experiment is that the edge region is metallic\cite{Kashiwaya2014}, with a superconducting gap developing only further into the sample.

Such a scenario is plausible given the fragility of unconventional superconductivity to elastic scattering (i.e. the inapplicability of Anderson's Theorem to a sign-changing order parameter), which has been explicitly verified for this material\cite{Mackenzie1998}.  Rough or pair-breaking surface effects have been shown\cite{Ashby2009,Nagato1998} to sharply reduce the superconducting order parameter at the surface, although not to meaningfully alter the surface density of states. Accordingly, the observation of metallic behavior suggests that there is a higher density of defects near the surface (presumably introduced during crystal growth or preparation procedures), leading to a reduced mean free path and the quenching of superconductivity near the surface.

To facilitate calculations, we do not directly treat a rough surface or defects in the surface region, but rather adopt a model consisting of a clean interface between vacuum and a metallic region, which in turn has a clean interface with the superconducting bulk.  The metallic region is arranged by setting appropriate coupling constants to zero in lattice Bogoliubov de-Gennes Hamiltonians. This introduces artifacts which will be discussed in section VII.

\section{Model Hamiltonians}
We consider spinless fermions on a 2D square lattice corresponding to the RuO$_2$ plane, and work in a cylinder geometry: periodic boundary conditions are taken in the $y$ direction, and open boundary conditions in $x$. We will consider two different Bogoliubov-de-Gennes Hamiltonians:
\begin{align}
&H_{\gamma}=-\sum_{i,j}T^z_{ij}c^{\dagger}_{z,i}c_{z,j} \nonumber \\
&+\sum_i\left[\Delta^{\gamma}_x(i)c^{\dagger}_{z,i}c^{\dagger}_{z,i+\hat x}+\Delta^{\gamma}_y(i)c^{\dagger}_{z,i}c^{\dagger}_{z,i+\hat y}+h.c.\right]
\end{align}

\begin{align}
H_{\alpha\beta}=&-\sum_{i,j}\sum_{\eta=x,y}T^{\eta}_{ij}c^{\dagger}_{\eta,i}c_{\eta,j}\nonumber \\
-&t'\sum_i\sum_{s=\pm 1} s \left[c^{\dagger}_{x,i}c_{y,i+ \hat x+s\hat y}+h.c.\right] \nonumber \\
&+\sum_i\sum_{s=\pm 1} \left[\Delta^{\alpha\beta}_x(i)c^{\dagger}_{x,i}c^{\dagger}_{x,i+\hat x+s\hat y}+ \right. \nonumber \\
&\left. s\Delta^{\alpha\beta}_y(i)c^{\dagger}_{y,i}c^{\dagger}_{y,i+\hat x+s\hat y}+h.c.\right]
\end{align}
$H_{\gamma}$ is a minimal Hamiltonian for chiral $p$-wave superconductivity on the $\gamma$ band of Sr$_2$RuO$_4$, which arises principally from Ru $4d$ $d_{xy}$ orbitals (represented by the index $z$ on fermion operators), for which we include the tight binding matrix elements $t_z\equiv T^z_{i,i\pm\hat x}=T^z_{i,i\pm\hat y} $, $t'_z\equiv T^z_{i,i\pm\hat x\pm \hat y} $, $\mu_z \equiv T^z_{i,i}$.  $H_{\alpha\beta}$ corresponds to the quasi-one-dimensional $\alpha$ and $\beta$ bands, which arise principally from the $d_{xz}$ and $d_{yz}$ orbitals (fermion indices $x$ and $y$ respectively), with tight binding matrix elements $t\equiv T^x_{i,i\pm\hat x}=T^y_{i,i\pm\hat y}$, $t_{\perp}\equiv T^x_{i,i\pm \hat y} = T^y_{i,i\pm \hat x}$, $\mu\equiv T^x_{i,i}=T^y_{i,i}$. For this model there is also an important next-nearest-neighbor orbital hybridization matrix element $t'$, whose presence is crucial for establishing a chiral superconducting gap. We take values $\{ t,t_{\perp},t',\mu,t_z,t'_z,\mu_z\}=\{1,0.1,0.1,1,0.8,0.3,1.15\}$ which are consistent with the Fermi surface measured in ARPES\cite{Damascelli2000} and the quasiparticle effective masses measured in quantum oscillations\cite{Bergemann2003}.

Nearest-neighbor pairing for the $d_{xy}$ orbital and next-nearest neighbor pairing for the $d_{xz}$ and $d_{yz}$ orbitals represent the lowest lattice harmonics consistent with a weak coupling analysis\cite{Raghu2010}, which predicts a fully gapped $d_{xy}$ orbital and "accidental" nodes on  $d_{xz}$ and $d_{yz}$ which are lifted to parametrically deep gap minima in the presence of orbital mixing $t'$. Calculations are performed with the self-consistency conditions $\Delta^{\gamma}_x(i)=-g_{\gamma}(i)\langle c_{z,i+\hat x}c_{z,i}\rangle$, $\Delta^{\gamma}_y(i)=-g_{\gamma}(i)\langle c_{z,i+\hat y}c_{z,i}\rangle$, $\Delta^{\alpha\beta}_x(i)=-g_{\alpha\beta}(i)\langle c_{x,i+\hat x +\hat y}c_{x,i}\rangle$, $\Delta^{\alpha\beta}_y(i)=-g_{\alpha\beta}(i)\langle c_{y,i+\hat x +\hat y}c_{y,i}\rangle$ with attractive interactions $g_{\alpha\beta}(i)$ and $g_{\gamma}(i)$ which are allowed to vary along the $x$ direction.  We model the metallic edge region adjoining the superconducting bulk by setting  $g_{\alpha\beta}$ and $g_{\gamma}$ to zero in a region of width $N_m$ sites, and nonzero and uniform in a region of width $N_s$ sites, with value chosen to yield the desired bulk values of $\Delta^{\alpha\beta}$ and $\Delta^{\gamma}$. In this model, superconductivity arises independently on the quasi-two-dimensional $\gamma$ band and on the quasi-one-dimensional $\alpha$ and $\beta$ bands (i.e. there is no inter-band proximity effect) and our estimate for the Sr$_2$RuO$_4$ edge current will be the sum of contributions from $H_{\gamma}$ and $H_{\alpha\beta}$. The consequences of this artificial assumption will be considered in section VII.

The current operator for the link from site $i$ to site $j$ can be derived from the lattice version of the equation of continuity and the Heisenberg equation of motion.  It has an intra-orbital part
\begin{equation}
\hat J^{\eta}_{i,j}=i  T^{\eta}_{i,j} \left[c^{\dagger}_{\eta,i,}c_{\eta,j}-h.c.\right]
\end{equation}
where $\eta=x,y,z$ is the orbital index. For the model of the $\alpha$ and $\beta$ bands there is also an inter-orbital part for the current between next-nearest neighbors
\begin{align}
\hat J^{xy}_{i,i+s_1\hat x+s_2\hat y}=&i  t' s_1s_2 \left[c^{\dagger}_{x,i,}c_{y,i+s_1\hat x +s_2\hat y}\right.+\nonumber\\ &\left.c^{\dagger}_{y,i,}c_{x,i+s_1\hat x +s_2\hat y} -h.c.\right]
\end{align}
where $s_1,s_2=\pm1$.

We neglect the effect of screening, whose effects have been explored elsewhere~\cite{Matsumoto1999,Furusaki2001,Ashby2009}.  Accordingly, our figure of merit for edge currents will be the total amount of current $I$ flowing through the metal region and half of the superconducting bulk, i.e. 
\begin{equation}
I=\sum^{N_m+N_s/2}_{n=1} \langle \hat J_{n\hat x, n\hat x+\hat y}+\hat J_{n\hat x, n\hat x+\hat x+ \hat y}\rangle
\end{equation}
where the two terms in the sum are for nearest neighbor and next-nearest neighbor links, including intra- and inter-orbital contributions as appropriate, and the angle brackets represent a thermal average.  Note that only net currents in the $\hat y$ direction are allowed by continuity in the cylinder geometry.
\section{Ginzburg Landau Theory}
Ginzburg-Landau theory represents an approximate solution to the BdG equations that becomes exact in the limit $T-T_c\rightarrow 0^-$, but provides valuable intuition even at low temperatures. The expression for the free energy can be found in the literature\cite{Sigrist1991}:
\begin{align}
F=&r\left(|\psi_x|^2+|\psi_y|^2 \right)+K_1\left(|\partial_x\psi_x|^2+|\partial_y\psi_y|^2\right)\nonumber \\
&+K_2\left(|\partial_y\psi_x|^2+|\partial_x\psi_y|^2\right)\nonumber \\
&+K_3\left([\partial_x \psi_x]^*[\partial_y\psi_y]+[\partial_y \psi_x]^*[\partial_x\psi_y]+c.c.\right)\nonumber \\
&+ \mathrm{higher} \quad \mathrm{order} \quad  \mathrm{terms}.
\end{align}
For our purposes, we need not treat quartic terms or those with more than two derivatives. The equations for the order parameter fields must be supplemented by appropriate conditions for a boundary at fixed $x$:
\begin{align}
\psi_x=0,\quad \partial_x\psi_y=0, \qquad \mathrm{insulating} \quad \mathrm{boundary} \\
\partial_x \psi_x = \frac{\psi_x}{b_x},\quad \partial_x \psi_y = \frac{\psi_y}{b_y}, \qquad \mathrm{metallic} \quad \mathrm{boundary}
\end{align}
The conditions for an insulating boundary follow from the fact that specular scattering is fully pair-breaking for $\psi_x$ (which is by construction odd under $x\rightarrow -x$)\cite{Ambegaokar1974}. The conditions for a metallic boundary involve phenomenological parameters $b_{x,y}$ which capture the fact that a metal interface is partially pair-breaking for both components \cite{DeGennes1991}.

We  continue to ignore screening, and focus on the spontaneous current (i.e. the current which exists in the absence of phase gradients imposed by an external field):
\begin{align}
\label{glcurrent}
J_{spont}&\propto -iK_3\left( \psi_y[\partial_x\psi_x]^*+\psi_x[\partial_x\psi_y]^*-c.c.\right) \nonumber \\
&\propto K_3\left(|\psi_y|\partial_x |\psi_x|- |\psi_x|\partial_x |\psi_y|\right)
\end{align}
In these expressions we have implemented translation symmetry in the $y$ direction and assumed a uniform relative phase factor of $i$ between $\psi_x$ and $\psi_y$ (i.e. positive chirality). Here the coefficients $K_1$ and $K_2$ determine the coherence lengths of the two order parameter components, the inter-component gradient coupling $K_3$ sets the scale of the currents, and $r\propto T-T_c$ is the usual parameter which tunes through the critical point.  The coefficients can be treated as phenomenological parameters or computed directly from the microscopic Hamiltonians given above.

\section{BdG Results}
\begin{figure}%
\centering
\subfloat[][]{\includegraphics[width=8cm, height=6cm]{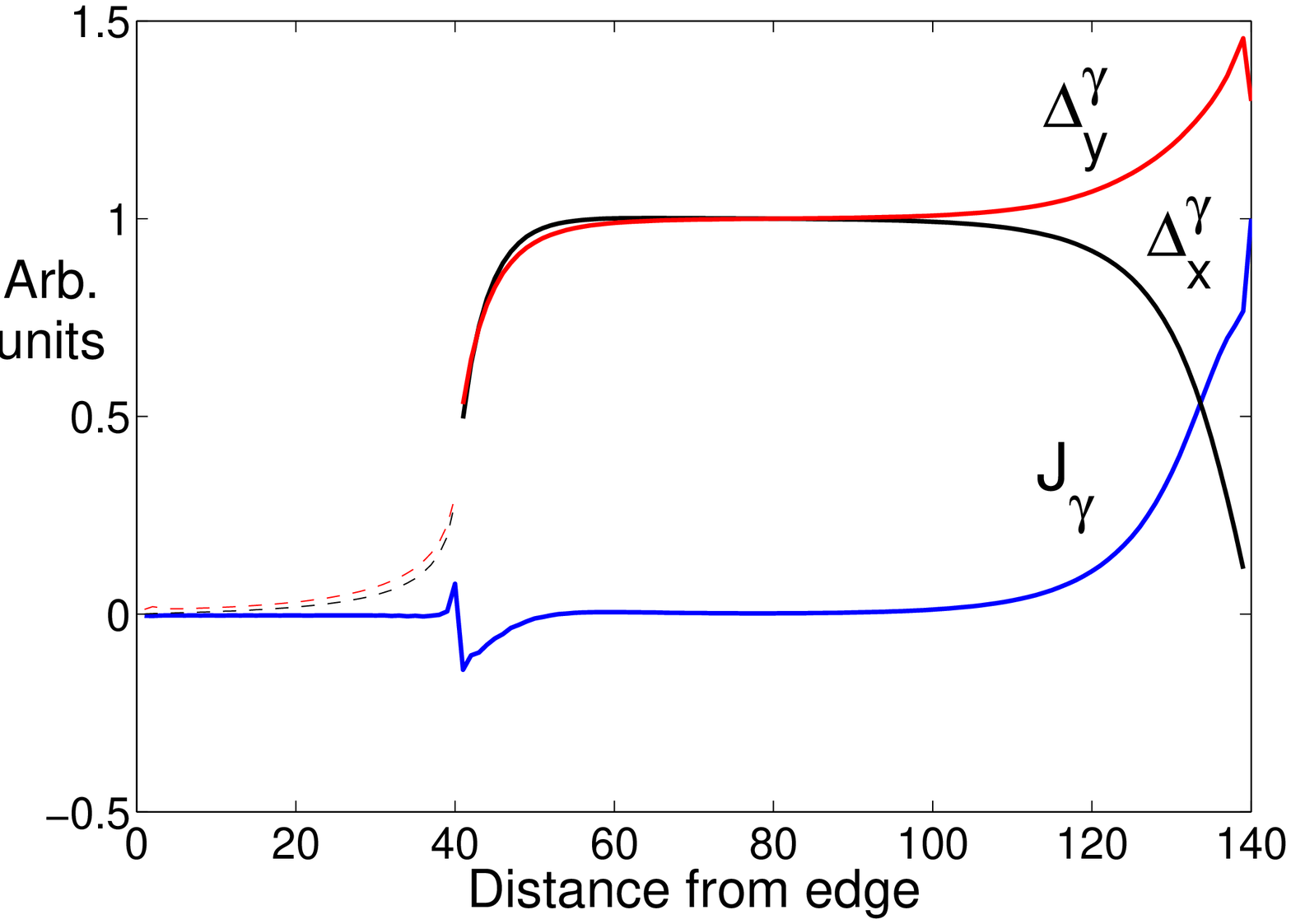}}%
\quad\subfloat[][]{\includegraphics[width=8cm, height=6cm]{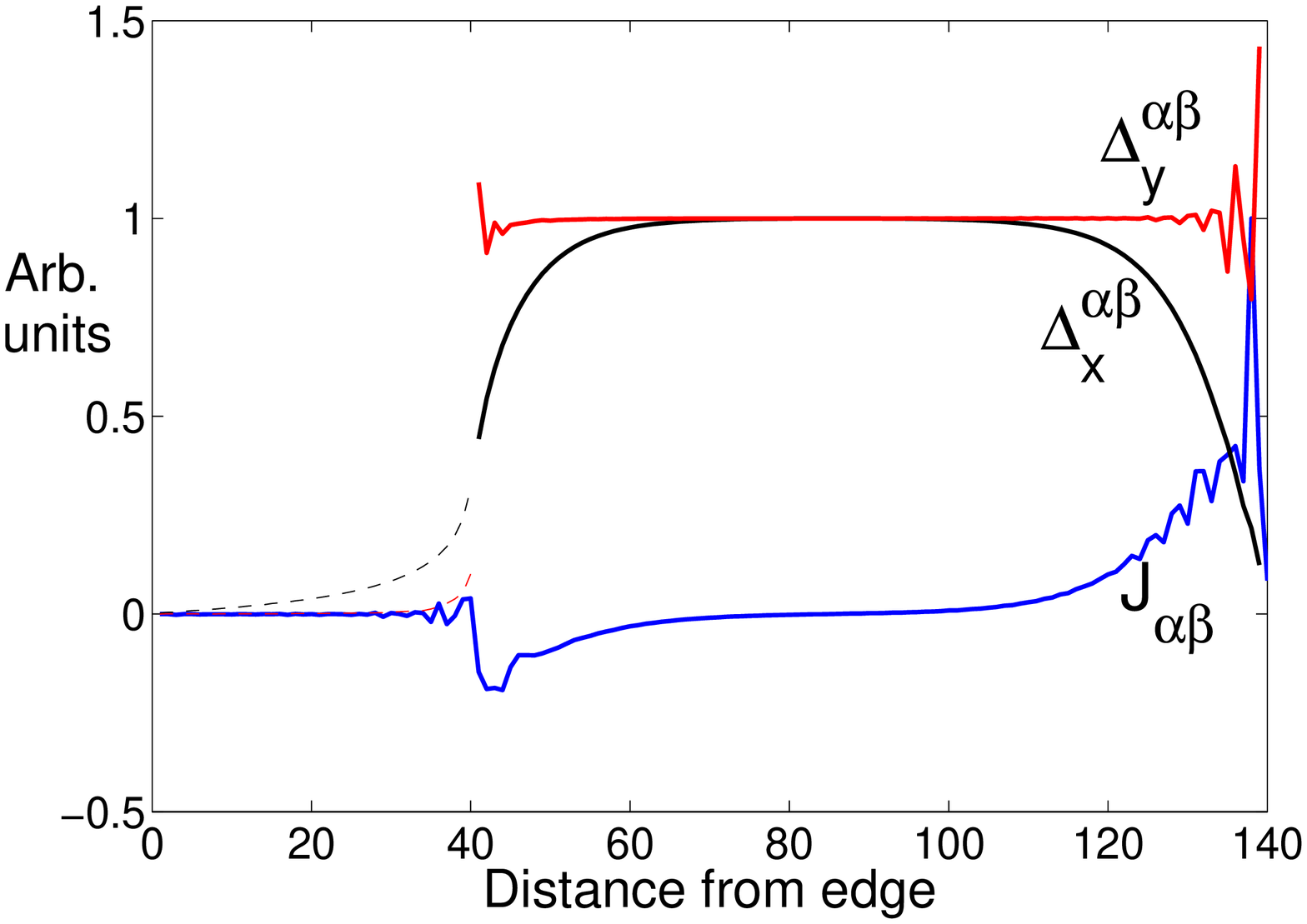}}
\caption{Current and the two components of the order parameter as a function of position for (a)$H_{\gamma}$ and (b) $H_{\alpha\beta}$. The first 40 sites are the metallic region, in which the gap vanishes, and clean interfaces with vacuum are present at positions $0$ and $140$. Pair correlations in the metallic region are shown in dashed lines. The bulk order parameter values are $\Delta^{\alpha\beta}_0=\Delta^{\gamma}_0=0.05t$, $T=0.2T_c$}.
\label{fig:cont}%
\end{figure}

\begin{figure}%
\centering
\subfloat[][]{\includegraphics[width=8cm, height=6cm]{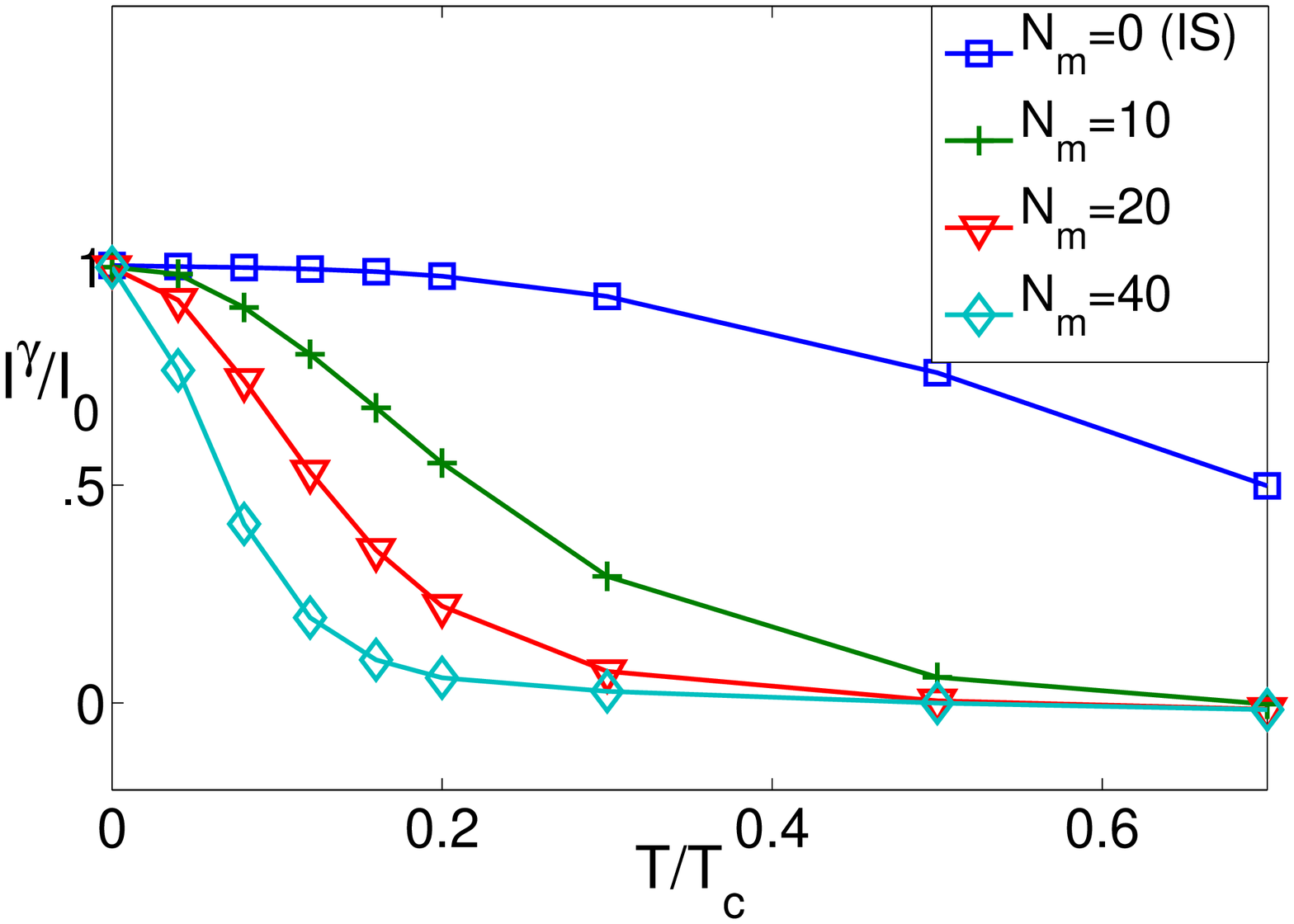}}%
\quad\subfloat[][]{\includegraphics[width=8cm, height=6cm]{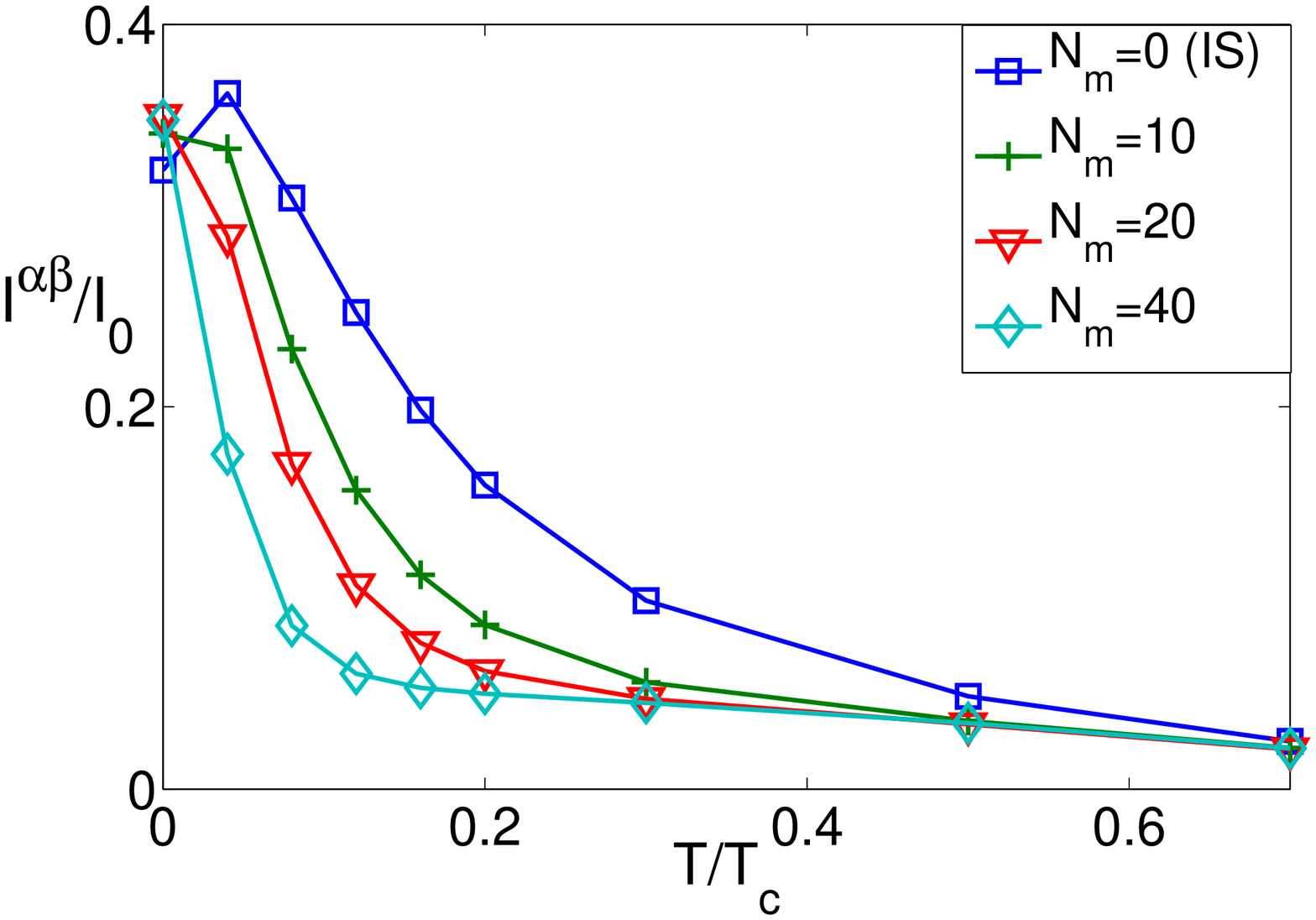}}
\caption{Contributions to the current near a metallic edge region from (a) the $\gamma$ band and (b) the $\alpha,\beta$ bands vs. temperature for several values of $N_m$, the thickness of this metallic edge region abutting the superconducting bulk. The superconducting bulk is of width $N_s=100$ sites,  and currents are quoted in units of $I_0$, which is essentially the Matsumoto-Sigrist result\cite{Matsumoto1999} in the absence of screening. The bulk order parameter values are $\Delta^{\alpha\beta}_0=\Delta^{\gamma}_0=0.05t$. For the current from $\alpha,\beta$ there are finite size effects associated with near-nodal quasiparticles which render the results at very low temperature less well behaved. We have verified that the zero-temperature current values in the thermodynamic limit are within 15$\%$ of those shown here.}\label{fig:cont}
\end{figure}

\begin{figure}%
\centering
\subfloat[][]{\includegraphics[width=8cm, height=6cm]{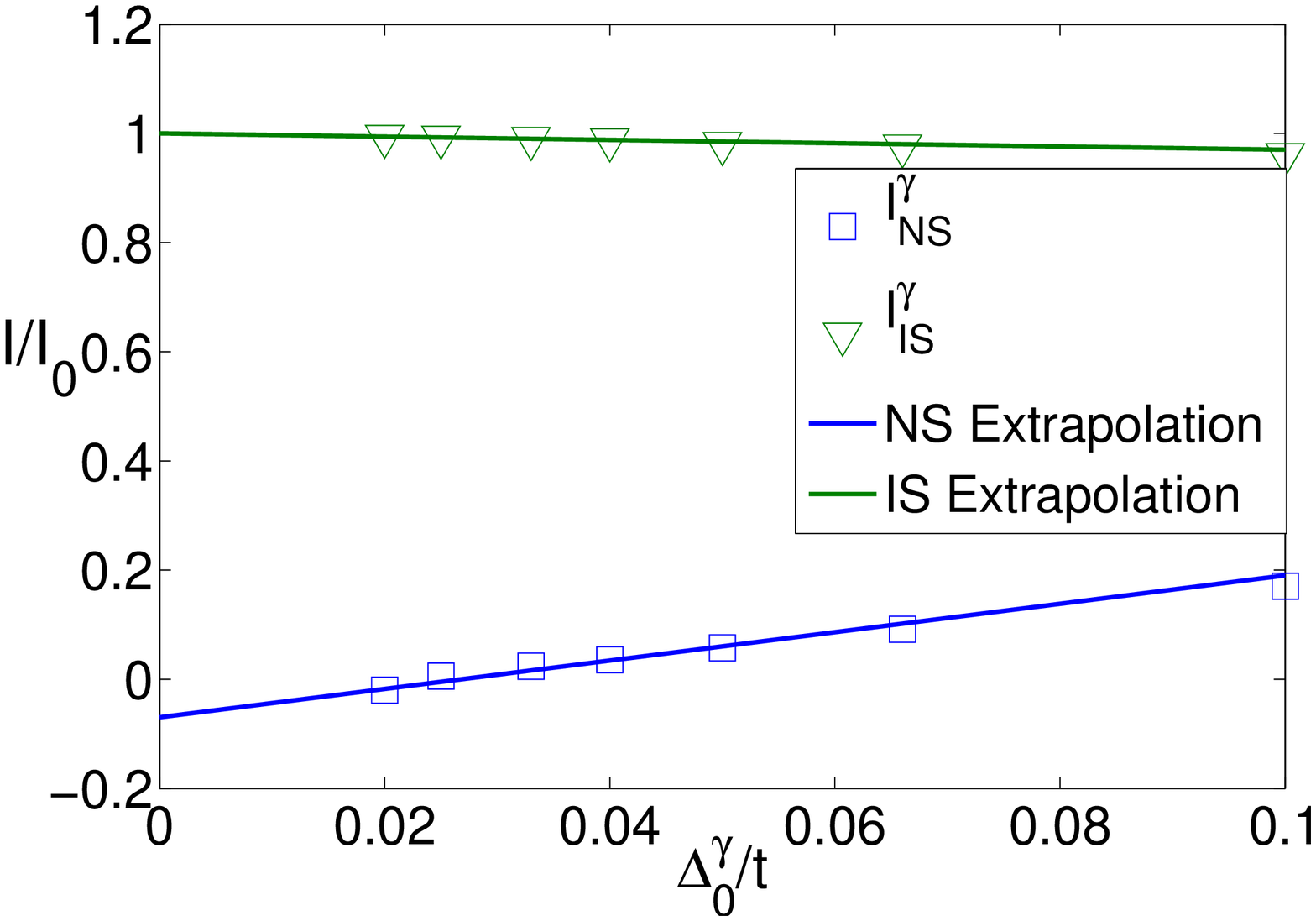}}%
\quad\subfloat[][]{\includegraphics[width=8cm, height=6cm]{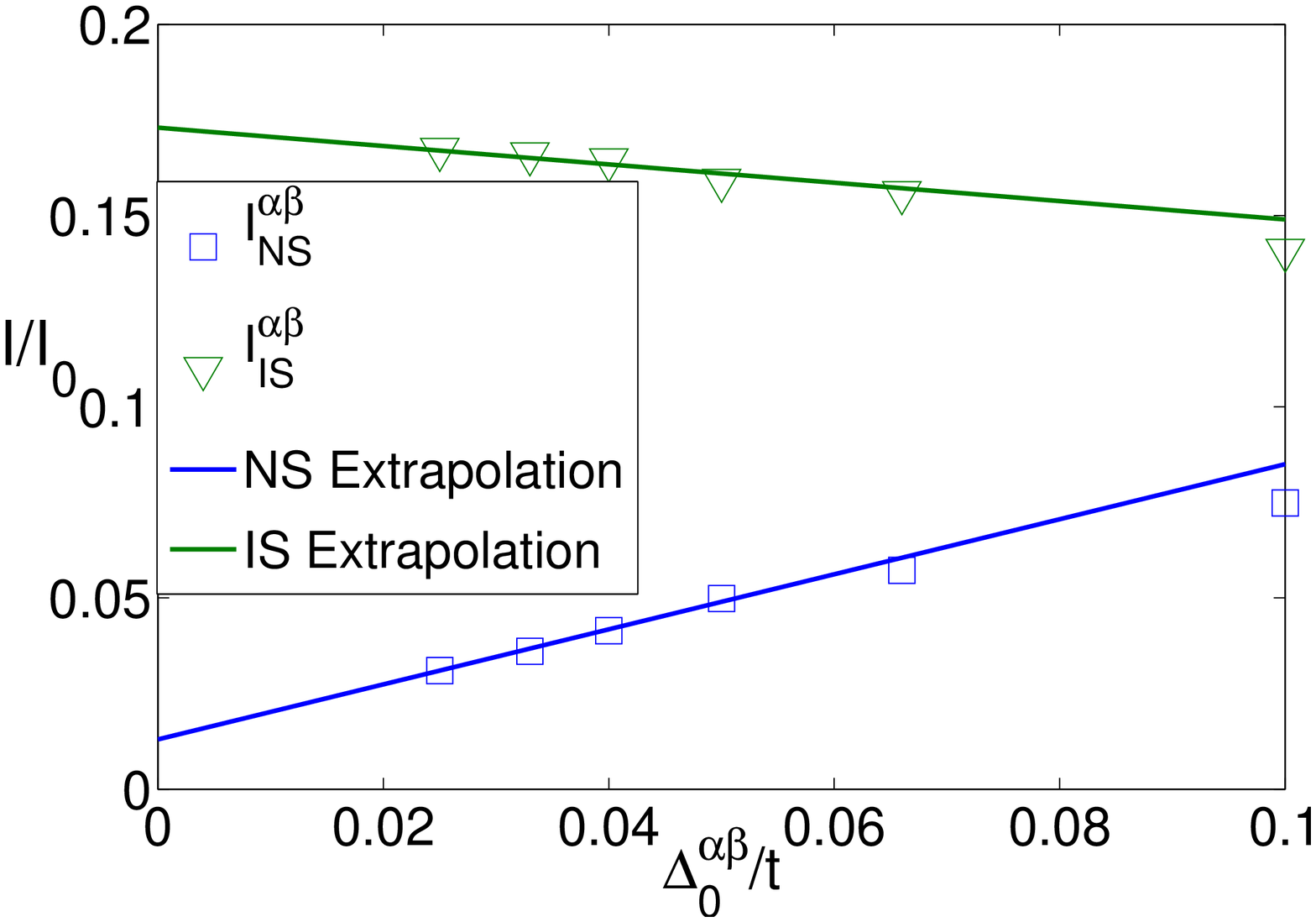}}
\caption{Extrapolation of current contributions of (a) the $\gamma$ band and (b) the $\alpha,\beta$ bands to the weak coupling limit $\Delta^{\alpha\beta}_0,\Delta^{\gamma}_0\rightarrow 0^+$ for the both insulator/superconductor (IS) and normal metal/superconductor (NS) interfaces. Currents are quoted in units of $I_0$ which is essentially the Matsumoto-Sigrist result\cite{Matsumoto1999} in the absence of screening. As the bulk gap is reduced, the temperature is reduced and the length scales $N_m$, $N_s$ are increased in order to fix the values $T/T_c=0.2$, $N_m/\xi\approx 4$, $N_s/\xi\approx 12$. The metallic boundary leads to suppression of over an order of magnitude in both the quasi-1D and quasi-2D cases. }
\end{figure}

As previously mentioned, our estimate for the edge current in Sr$_2$RuO$_4$ is the sum of contributions due to the quasi-2D $\gamma$ band and the quasi-1D $\alpha, \beta$ bands; we initially plot and discuss these contributions separately. Values of net current are given in units of $I_0\equiv 0.073  et/\hbar$, which is the net current due to the $\gamma$ band with an insulating interface ($N_m=0$) at $T=0.2 T_c$, in the weak coupling limit $\Delta^{\gamma}_0\rightarrow0^+$.  $I_0$ is approximately equal to the value of the total current per spin in a quasi-classical approximation (such as the Matsumoto-Sigrist prediction\cite{Matsumoto1999} used in \cite{Kirtley2007,Hicks2010}) when screening is neglected. If our model predicts a current $I$ and screening alters our predictions in the same way as it does the quasi-classical results of Matsumoto-Sigrist, then our prediction of a magnetic signal (such as the peak flux) is equal to the Matsumoto-Sigrist prediction times $I/I_0$.

Plots of the current and both components of the order parameter as a function of distance from the edge are shown in Figures 1. Figures 2 show the two current contributions versus temperature for several choices of $N_m$. Data points near $T_c$ are not included due to computational cost.  Figures 3 show the current contributions as a function of the bulk order parameter ($\Delta^{\gamma}$ and $\Delta^{\alpha\beta}$ respectively, with fixed values of $T/T_c$ and $N_m / \xi_0$. Before considering the effect of the normal-metal region, we note basic results for a clean insulator (or vacuum) / superconductor (IS) interface ($N_m=0$). In that case, compared to the contribution from $H_{\gamma}$, the net current from $H_{\alpha\beta}$ is reduced by a factor of approximately three at zero temperature and six at the experimental temperature of $0.2 T_c$.  

Turning to the results for a normal metal / superconductor (NS) interface (i.e. $N_m\neq 0$), one feature of the $I-T$ curves for different values of $N_m$ is that they all coincide at zero temperature and at sufficiently high temperature, differing only in an intermediate crossover region.  This follows from the proximity effect: while the superconducting gap $\Delta(i)\equiv -g \langle c c \rangle$ is zero in the metal (where $g=0$), pair correlations $\langle cc \rangle$ do penetrate.  The length scale for this penetration is set by $v_F/T$, (where $v_F$ is the Fermi velocity), and thus diverges at zero temperature, so that the width of the metallic region is effectively zero.  By contrast, at temperatures such that $v_F/T <N_m$, pairing correlations decay to zero before the edge is encountered, so that the metallic region is effectively infinite.  In both cases, an increase in $N_m$ should have a negligible effect on the currents, consistent with the calculation.

For $v_F/T <N_m$ there is a pronounced suppression of the current in both the one and quasi-1D cases compared with the current without a metallic region ($N_m=0$). The amount of this suppression depends on the size of the pairing gap. For Sr$_2$RuO$_4$, the pairing gap is on the order of $10^{-3} t$, so that extrapolation to the weak coupling limit $\Delta_0\rightarrow 0^+$ is necessary for a quantitative estimate. For a model including all three bands in this weak coupling limit, we find a suppression of approximately twenty compared to the initial Matsumoto-Sigrist predictions.
\section{Qualitative explanation from Ginzburg-Landau theory}
The results of the previous section can be summarized as follows: 1) the contribution from the $\alpha,\beta$ bands is a several times smaller than that of the $\gamma$ band for the IS geometry. 2) both contributions are substantially suppressed in the NS geometry. 3) the suppression due to the NS geometry is considerably larger for the $\gamma$ band than for the $\alpha,\beta$ bands. Ginzburg-Landau theory, though it is not quantitatively valid at low temperatures, can nonetheless qualitatively explain each of these results.

1) With a conventional insulating interface, the scale of spontaneous currents is set by the coefficient $K_3$. In the quasi-2D model, this is a number of order one, whereas in the quasi-1D model, it vanishes in the limit of zero inter-orbital mixing $t'$. Since $t' = 0.1t$, it follows that $K_3^{\alpha\beta}$ is substantially smaller than  $K_3^{\gamma}$ and similarly for the currents. A microscopic calculation gives $K_3^{\alpha\beta}\approx 0.02 K_3^{\gamma}$.

2) The suppression in current in the NS geometry can be viewed as a consequence of the different boundary conditions on the order parameter. The boundary values of of $|\psi_x|$ and $|\psi_y|$ are respectively increased and decreased compared to the insulating case. At a fixed distance from the edge $|\psi_x|$ and $\partial_x |\psi_y|$ are larger while $|\psi_y|$ and $\partial_x |\psi_x|$ are smaller than their corresponding values for the insulating boundary. Eq. \eqref{glcurrent} for the current shows that this yields a numerical (though not parametric) reduction in the current for any choice of G-L coefficients.

3) The tremendous suppression of the current in the quasi-2D NS model is a lattice effect.  For the fine-tuned case $K_1=K_2$, one can show that $b_x=b_y$ and the two components of the order parameter heal away from the metal in precisely the same way, leading to a vanishing current in lowest-order G-L theory\cite{Ashby2009}. For a quadratic dispersion and an order parameter $k_x+ik_y$, as is often used to describe the $\gamma$ band\cite{Matsumoto1999,Sauls2011,Furusaki2001,Stone2004}, the coefficients satisfy $K_1=3K_2$. However, for a lattice-compatible order parameter $\sin k_x + i \sin k_y$ as treated here and for an appropriate tight-binding band structure for the $\gamma$ band, $K_1=0.71K_2$.  The large suppression of the $\gamma$ band current due to the NS geometry can be roughly identified with the proximity of this result to the fine-tuned case $K_1=K_2$.

\section{Discussion}
Superconductivity on the quasi-1D bands was previously conjectured\cite{Raghu2010} to lead to dramatically reduced edge currents compared to a quasi-2D scenario due to trivial topology (i.e. the Chern numbers of the two bands add to zero, yielding no net chiral edge modes). The results shown above for the IS interface show a substantial reduction (by a factor between three and six), but nonetheless of order one, falsifying the initial conjecture and illustrating the tenuous connection between topology and edge currents in chiral $p$-wave superconductors (this topic will be treated in depth in a forthcoming paper). 

Even if the quasi-1D bands had vastly reduced currents in the IS case, the contribution from the $\gamma$ band would generically be large, even if it were not the "dominant" band.  The neglect of the current contribution from the subdominant band(s) is only justified if the experimental temperature exceeds the subdominant gap scale.  However, thermodynamic evidence shows that the gaps on all bands are at least comparable to $T_c = 1.5 K \approx .13 meV$\cite{Firmo2013}.  At low temperatures, the edge currents should then correspond to the sum of contributions from the quasi-1D and quasi-2D bands, with the weak coupling limit taken for both $\Delta^{\alpha\beta}$ and $\Delta^{\gamma}$.  At low temperatures and with a clean interface, the generic scale of edge currents is "of order one" regardless of microscopic mechanism details such as the identity of the dominant band(s).

Though there does not seem to be any physical reason for a parametric suppression of edge currents, we find a meaningful quantitative reduction of over an order of magnitude compared to previous estimates by considering the effect of surface imperfection. Within a model of a clean metal of width $\sim 4 \xi_0$ abutting a clean superconductor, with $T=0.2 T_c$, the total current from all three bands is suppressed by a factor of more than twenty in the weak coupling limit compared to the result for the $\gamma$ band and an IS interface. Within our model, there is essentially no suppression in the limit of sufficiently low temperatures and/or narrow metallic regions, where superconducting correlations induced by the proximity effect extend all the way to the edge.  This is an artifact of our model, however, which does not treat surface roughness or disorder directly.   For example, pair-breaking and diffuse scattering effects are known to reduce the zero-temperature current \cite{Ashby2009,Nagato1998}. 

The calculations presented here are not expected to be quantitatively correct for the actual superconducting gap structure and surface physics of Sr$_2$RuO$_4$.  Our model of spinless fermions entirely neglects spin-orbit coupling (SOC), which has been proposed to qualitatively affect pairing\cite{Veenstra2014}.  However, as far as the edge current is concerned, the primary effect of SOC is to modestly renormalize the band structure; hence, its explicit inclusion would not change any of our results substantially.  A more serious unphysical assumption is the neglect of the inter-band proximity effect, without which superconductivity would generically arise at very different temperatures on the $\gamma$ and $\alpha/\beta$ bands.  While inter-band proximity coupling would not change the additivity of the current contributions from the different bands, it would alter the length scale over which the various order parameter components heal away from an interface. The resulting currents could be reduced or increased compared to our results, depending on microscopic details.  

These defects notwithstanding, the model treated above illustrates that substantial reductions in magnetic signal can arise from interface effects. We now consider the consequences of a twenty-fold reduction for the interpretation of magnetometry experiments. Even with this reduction the magnetic signal at the edge would still be estimated to be several times the resolution of scanning magnetometry experiments, and should therefore be observable. However, if multiple domains of sufficiently small size are present in the sample and intersect the edge, the magnetic fields from spontaneous currents would be unobservable.  Kirtley et al\cite{Kirtley2007} find that, to be consistent with the Matsumoto-Sigrist predictions\cite{Matsumoto1999}, $ab$-plane domains below about $1.5 \mu m$ in size are necessary.  To be consistent with a prediction twenty times smaller, the domains could be as large as perhaps $5 \mu m$.  However, the presence of multiple $ab$-plane domains within the sample would lead to spontaneous currents at the domain walls, which have not been treated here.  Unless domain walls are pinned by crystal defects that, like a rough edge, lead to quenched superconductivity (an unlikely proposition), the suppression indicated in the foregoing calculations would not apply to the domain wall currents.  

One scenario for the lack of an edge signal which would not imply a signal at interior domain boundaries is the c-axis stacking of planar domains of macroscopic horizontal extent and alternating chirality.  The energetic cost of the domain boundaries would be small, due to the very weak dispersion of the electronic band structure along the c direction, and symmetry requires that no spontaneous current would flow at these boundaries.  The measurements of Hicks et al \cite{Hicks2010} place an upper bound of $20-400nm$ on the height of such domains (depending on microscopic domain details, and again assuming Matsumoto-Sigrist predictions for edge currents\cite{Matsumoto1999}).  Here, a twenty-fold reduction of expected edge currents for a single domain would revise upward the experimental bound on domain size, possibly reconciling the null result of scanning magnetometry experiments with the spontaneous time reversal symmetry breaking seen in Kerr effect measurements with mesoscopic spot size ($\sim 50 \mu m$) and skin depth ($\sim 150 nm $\cite{Kapitulnik2014}).

We have shown that spontaneous currents in a chiral $p$-wave superconductor are highly sensitive to interface details, in particular that surface disorder leading to a $~\mu m$-thickness metallic surface region can cause a suppression of more than an order of magnitude compared to naive estimates.  We propose that a scenario of c-axis domain stacking, along with surface disorder, might resolve the seeming disagreement between scanning magnetometry and Kerr probes, and further suggest that the edge of a crystal fractured in vacuum might host a much lower defect density, and potentially lead to observable edge currents.

{\bf Acknowledgements:} SL thanks Aharon Kapitulnik, Steven Kivelson, Kathryn Moler, and Boris Spivak for helpful discussions.  This work is supported by NSERC and CIFAR at McMaster and by the Canada Research Chair and Canada Council Killam programs (CK). At Stanford, this work is supported in part by the DOE Office of Basic Energy Sciences, contract DE-AC02-76SF00515 (SL and SR), an ABB fellowship (SL), and the Alfred P. Sloan Foundation (SR).

\bibstyle{plain}
\bibliography{}

\end{document}